\documentclass[runningheads,a4paper]{llncs}

\usepackage{amssymb,amsmath}
\usepackage{graphicx}

\usepackage{url}
\urldef{\mailsa}\path{naldi@disp.uniroma2.it}
\newcommand{\keywords}[1]{\par\addvspace\baselineskip
\noindent\keywordname\enspace\ignorespaces#1}

\begin{document}

\mainmatter  

\title{Approximation of the truncated Zeta distribution and Zipf's law}

\titlerunning{Approximate Zeta distribution}

%
%
\author{Maurizio Naldi}
\authorrunning{M. Naldi}

\institute{Universit\`{a} di Roma Tor Vergata\\Department of Computer Science and Civil Engineering\\
Via del Politecnico 1, 00133 Roma, Italy\\
\mailsa\\
}

%
%

\toctitle{Lecture Notes in Computer Science}
\tocauthor{Authors' Instructions}
\maketitle

\begin{abstract}
Zipf's law appears in many application areas but does not have a closed form expression, which may make its use cumbersome. Since it coincides with the truncated version of the Zeta distribution, in this paper we propose three approximate closed form expressions for the truncated Zeta distribution, which may be employed for Zipf's law as well. The three approximations are based on the replacement of the sum occurring in Zipf's law with an integral, and are named respectively the integral approximation, the average integral approximation, and the trapezoidal approximation. While the first one is shown to be of little use, the trapezoidal approximation exhibits an error which is typically lower than 1\%, but is as low as 0.1\% for the range of values of the Zipf parameter below 1. 
\end{abstract}

\section{Introduction}
The Riemann Zeta function is a well known mathematical object with applications in many fields. For example, it is widely used in modern number theory; a number of its mathematical applications are shown in \cite{zagier1994values}. Among many other examples of its applications we  can cite quantum chaos theory \cite{Berry86} and cosmology \cite{Elizalde}. Two comprehensive surveys of the Riemann Zeta function are  contained in \cite{titchmarsh1986theory} and \cite{ivic2003riemann}, while a brief account of its characteristics can be read in \cite{hejhala2012alan}.

In many instances, the truncated version of the Riemann Zeta function is of interest, where the sum is limited to the first $n$ terms. For example, this truncated version arises in connection with the Zipf distribution (which is equivalent to the truncated Zeta distribution). Since the sum of the Zeta function has no closed expression, this reverberates on the Zipf distribution, which has no closed form.	

In this paper we provide three approximate expressions for the truncated Riemann Zeta function, named respectively integral approximation, average integral approximation, and trapezoidal approximation. We incorporate them in the truncated Zeta distribution and obtain closed form expressions for the individual terms of the Zipf distribution. We show that the accuracy of the integral approximation is acceptable just for very small values of the Zipf parameter (lower than 0.6), while the accuracy improves with the average integral approximation (whose error is lower than 10\% even when the Zipf parameter is 2). The best accuracy is achieved by the trapezoidal approximation, which exhibits an error that can be as low as 0.1\%. The focus is on expressions for values of the Zipf parameter different from 1.

The structure of the paper is as follows. In Section \ref{sec:zeta}, we review the basic notions about the Zeta distribution and Zipf's law. We then propose and examine the two approximations separately in Sections \ref{sec:integral} and \ref{sec:trapezio} respectively.

\section{The Zeta distribution and Zipf's law}
\label{sec:zeta}

The Zeta distribution is related to Zipf's law. In this section we review some basic characteristics of that distribution and show its relationship to Zipf's law.

The Zeta distribution (also named as the Joos model or the discrete Pareto distribution) has the following probability mass function (see \cite{Kotz-zeta} and \cite{johnson1970discrete} for definitions and basic properties, and \cite{zornig1995unified} for related distributions):
\begin{equation}
\label{zetainf}
\mathbb{P}[X=r]=\frac{r^{-\alpha}}{\zeta (\alpha)} \qquad r=1,2,\ldots \quad \alpha >0 ,
\end{equation}
where $\zeta (\alpha)$ is the Riemann Zeta function, tabulated for integer values of $\alpha$ (up to $\alpha = 42$) in \cite{Abra}. The expected value of $X$ exists just if $\alpha >2$ and is 
\begin{equation}
\label{mediazeta}
\mathbb{E}[X] = \frac{\zeta (\alpha -1)}{\zeta (\alpha)}.
\end{equation}
Similarly, the variance of $X$ exists just if $\alpha >3$ and is 
\begin{equation}
\label{varzeta}
\mathbb{V}[X] = \frac{\zeta (\alpha - 2)}{\zeta (\alpha)} - \left[  \frac{\zeta (\alpha -1)}{\zeta (\alpha)}\right]^2
\end{equation}
In general, the $m$-th moment of $X$ exists just if $\alpha > m+1$.

In many cases the variable $X$ can take only finite values. This bypasses the above limitations concerning the existence of its moments, since in this case its moments are always finite. A truncated version of the Zeta distribution is therefore of interest and can be written, by employing the partial sum of the Riemann Zeta function (which in turn allows us to remove the restriction on the value of $\alpha$), as:
\begin{equation}
\label{zetafin}
\mathbb{P}[X=r]=\frac{r^{-\alpha}}{\sum_{i=1}^{n}i^{-\alpha}} \quad \alpha >0.
\end{equation}

If we interpret $r$ as a rank variable and consider a population of individuals that distribute themselves among $n$ species, the model (\ref{zetafin}) can serve to represent the relative frequency of each species when the frequency-rank distribution obeys the generalized Zipf's law, after \cite{Zipf}. Zipf's law states that the product of a power of the rank $r$ and the frequency $f_{r}$ of the species of rank $r$ is constant for all ranks \cite{Read89}:
\begin{equation}
r^{\alpha}f_{r} = \textrm{const}.
\end{equation}
Zipf's law appears in many contexts: linguistics (frequency of words and of surnames) \cite{li1992random}, geostatistics (population of towns) \cite{maniadakis2010population}, Internet traffic (web access statistics), economics (distribution of incomes, size distribution of firms, market economics) \cite{Naldi2003,delli2008emergent,naldi2014interval,alegria2008measuring}, finance \cite{balakrishnan2008power}, service demand models \cite{colladon2013quality}, web click behaviour \cite{ali2007robust,Balakrishnan08,Regelson,Regelson,grillo2010penalized,naldi2010value}, teletraffic distribution \cite{Naldi2006,NaldiCN10,conti2012estimation}. The meaning attached to individuals and species depends on the particular application; e.g. for Internet traffic the species may be the websites and the individuals may be the single visits to each site.
The values for $n$ (the number of species) vary widely across the range of applications, from a few hundreds to tens of thousand, while $\alpha$ (in the following referred to as the Zipf parameter) is typically not too far from 1 (which would make the application of the Zeta distribution model impossible).

Despite its wide diffusion, no explicit form of (\ref{zetafin}) is found in the literature, due to the presence of the truncated Riemann Zeta function, while such a form would be useful in many studies involving this distribution. An approximation of the Riemann Zeta function is available only for the special case $\alpha =1$ (when this function is renamed the harmonic series), for which we have (see expr. 0.131 of \cite{Gradshteyn}):
\begin{equation}
\sum_{i=1}^{n}\frac{1}{i} \simeq \gamma + \ln (n) + \frac{1}{2n},
\end{equation}
where $\gamma$ is Euler's constant.

\section{The integral approximation}
\label{sec:integral}
After having seen that a basic hurdle for the derivation of a closed form expression for the truncated Zeta distribution is the lack of a closed form expression for the truncated Riemann Zeta function, in this section we provide an approximate expression, obtained by replacing  the sum in Equation (\ref{zetafin}) by the corresponding integral. In the following, we will refer to it as the integral approximation.

This replacement, as described in \cite{Graham1994}, provides the following approximation for the truncated Zeta function:
\begin{equation}
\label{appsum}
\sum_{i=1}^{n} i^{-\alpha} \simeq \int_{i}^{n+1}x^{-\alpha}d\alpha = \frac{(n+1)^{1-\alpha}-1}{1-\alpha},
\end{equation}
under the condition $\alpha \neq 1$. By replacing this expression in the Zeta distribution of Equation (\ref{zetafin}) we have
\begin{equation}
\label{appint}
\mathbb{P}[X=r] \simeq \mathbb{P}_{\textrm{int}}[X=r] = \frac{(1-\alpha)r^{-\alpha}}{(n+1)^{1-\alpha}-1}.
\end{equation}

\begin{figure}
\begin{center}
\includegraphics[width=0.7\columnwidth]{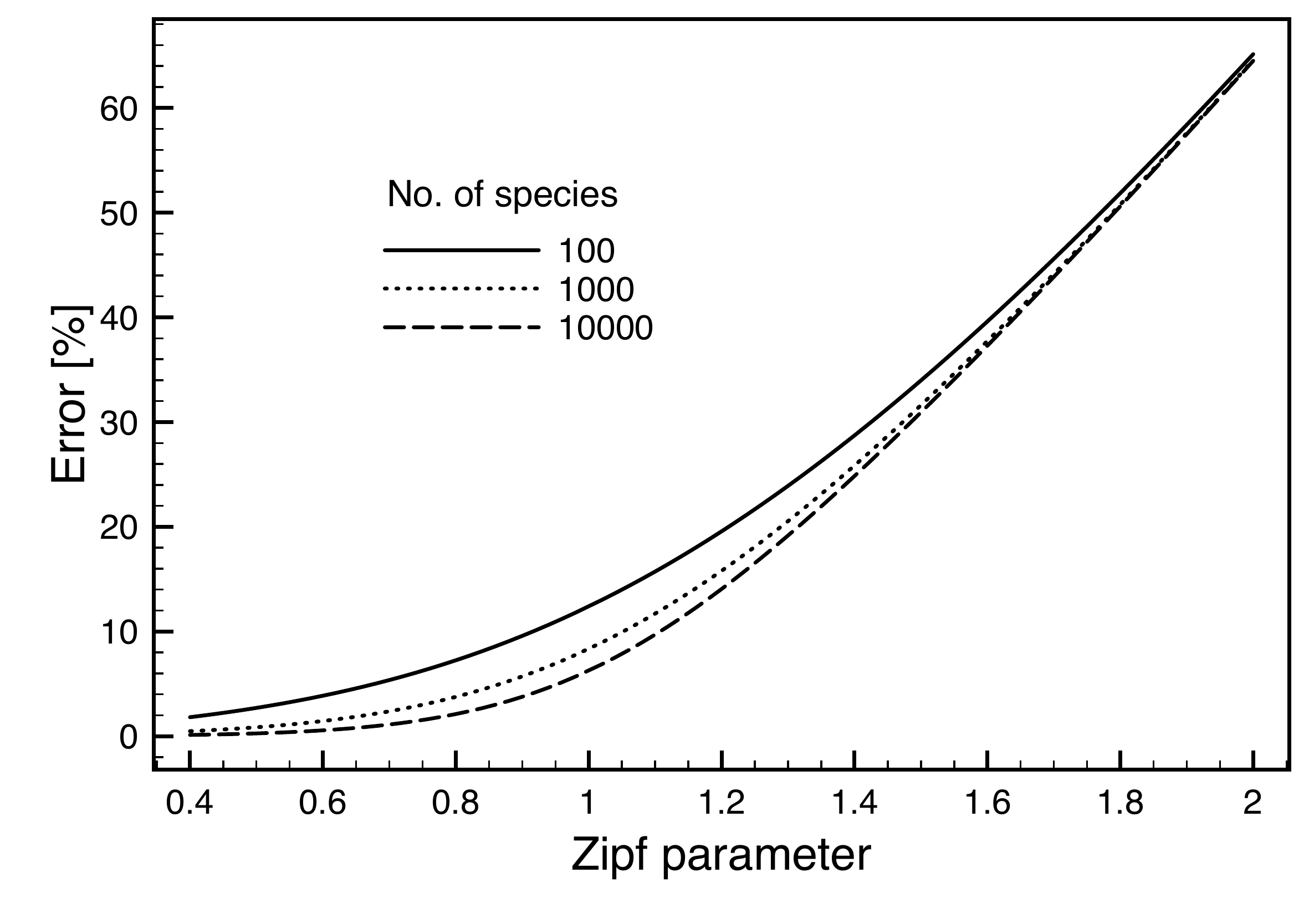} 
\caption{Relative error for the single integral approximation of the Zeta distribution}
\label{fig:errint}
\end{center}\end{figure}

The relative error we get when using this approximate expression instead of the exact one is
\begin{equation}
\label{err1}
\epsilon _{\textrm{int}} = \frac{\mathbb{P}_{\textrm{int}}[X=r]-\mathbb{P}[X=r]}{\mathbb{P}[X=r]} = \frac{(1-\alpha)\sum_{i=1}^{n}i^{-\alpha}}{(n+1)^{1-\alpha}-1}-1.
\end{equation}
It can be seen that it doesn't depend  on the rank $r$ and is plotted in \figurename~\ref{fig:errint} for three different values of the number of species ($n=100, 1000, 10000$).  This error grows very rapidly with $\alpha$ (but diminishes as the number of species grows) and makes the single integral approximation of little use except for the smallest values of the Zipf parameter.

However, the approximate expression (\ref{appsum}) is always smaller than the actual sum: the integrated curve lies under the staircase representing the series. We can instead think of an approximation based on the same principle of using the integral instead of the sum, but approximating it from the above, i.e.
\begin{equation}
\label{appabove}
\sum_{i=1}^{n}i^{-\alpha} \simeq 1+\int_{2}^{n+1}(x-1)^{-\alpha}dx = 1+\frac{n^{1-\alpha}-1}{1-\alpha}.
\end{equation}

By averaging the two expressions (\ref{appsum}) and (\ref{appabove}), we should obtain a more accurate expression than the single integral approximation:
\begin{equation}
\label{appave}
\begin{split}
\sum_{i=1}^{n}i^{-\alpha} &\simeq \frac{1}{2}\left[ \frac{(n+1)^{1-\alpha}-1}{1-\alpha} + 1+\frac{n^{1-\alpha}-1}{1-\alpha}\right]\\ &=\frac{(n+1)^{1-\alpha}+n^{1-\alpha}-(1+\alpha)}{2(1-\alpha)}.
\end{split}
\end{equation}
Under this average integral approximation, we have then a possibly refined integral approximation (which we call the average integral approximation) for the Zeta distribution
\begin{equation}
\label{zetaave}
\mathbb{P}_{\textrm{ave}}[X=r]=\frac{2(1-\alpha)r^{-\alpha}}{(n+1)^{1-\alpha}+n^{1-\alpha}-(1+\alpha)}.
\end{equation}

\begin{figure}
\begin{center}
\includegraphics[width=0.7\columnwidth]{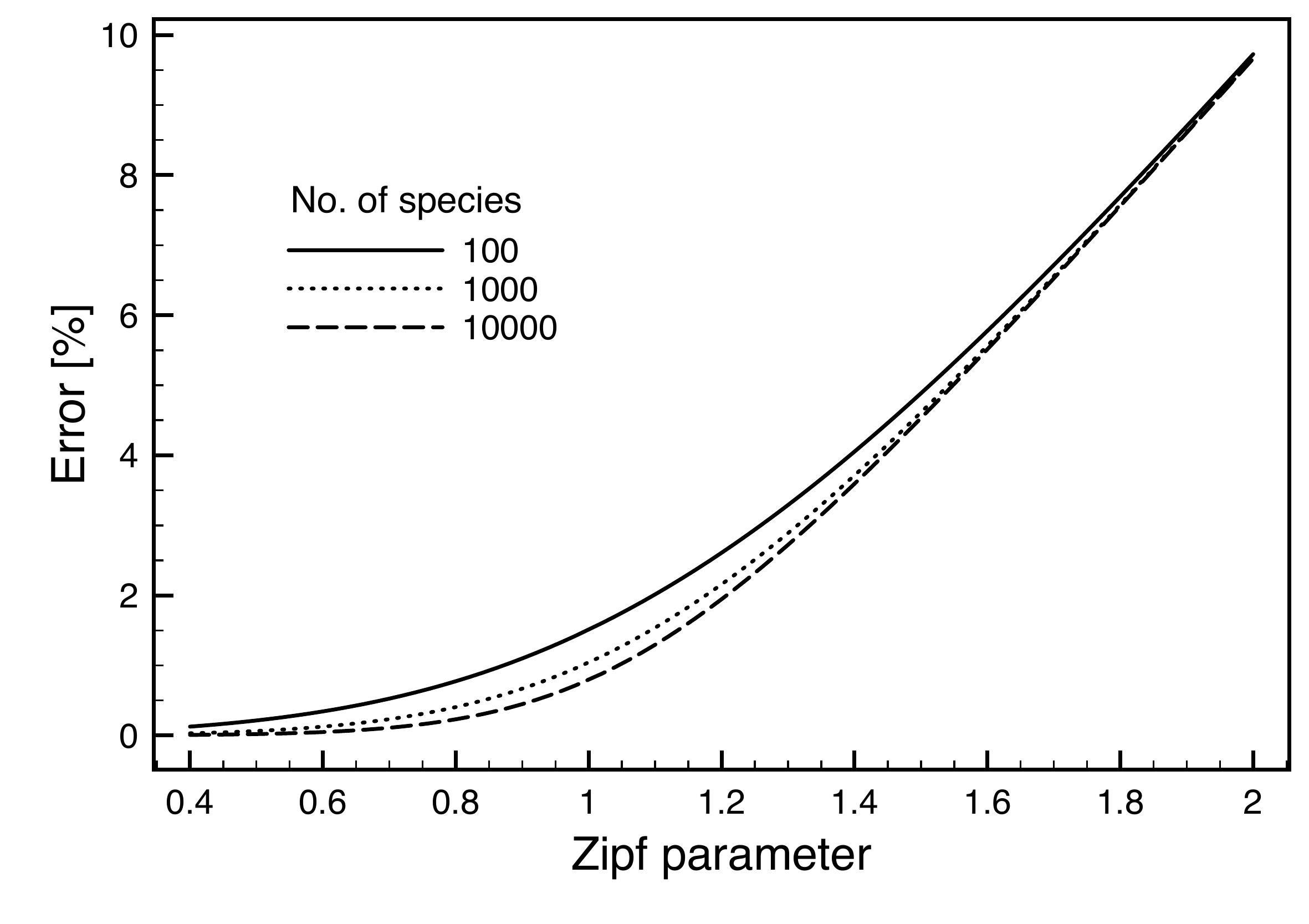} 
\caption{Relative error for the average integral approximation of the Zeta distribution}
\label{fig:errave}
\end{center}
\end{figure}

As for the single integral approximation, we can obtain the relative error
\begin{equation}
\label{err2}
\epsilon _{\textrm{ave}} = \frac{\mathbb{P}_{\textrm{ave}}[X=r]-\mathbb{P}[X=r]}{\mathbb{P}[X=r]} = \frac{2(1-\alpha)\sum_{i=1}^{n}i^{-\alpha}}{(n+1)^{1-\alpha}+n^{1-\alpha}-(1+\alpha)}-1.
\end{equation}
The resulting relative error is again independent of the rank $r$. As we can see in \figurename~\ref{fig:errave}, though exhibiting a fast growth with the parameter $\alpha$, it is much more controlled than the single integral approximation and stays below 10\%.

\section{The trapezoidal rule approximation}
\label{sec:trapezio}
Though providing a substantial improvement over the single integral approximation, the average integral approximation is still far from being satisfying especially for the largest values of the Zipf parameter $\alpha$. In this section we consider a different way of approximating the truncated Riemann Zeta function (and therefore the truncated Zeta distribution).

Instead of starting from the partial sum and approximating it with the corresponding integral, we start with this integral and approximate it through the trapezoidal rule, having the desired partial sum appear in the process.
Following this method, described by Kreminski \cite{kreminski1997using}, we have
\begin{equation}
\begin{split}
\int_{k}^{n}x^{-\alpha}dx &\simeq \sum_{i=k}^{n-1}
\frac{i^{-\alpha}+(i+1)^{-\alpha}}{2} = \frac{k^{-\alpha}}{2}
 + \frac{n^{-\alpha}}{2} +
 \sum_{i=k+1}^{n-1}i^{-\alpha} \\
&=\frac{k^{-\alpha}+n^{-\alpha}}{2}
+\sum_{i=1}^{n}i^{-\alpha}-\sum_{i=1}^{k}i^{-\alpha}
-n^{-\alpha}\\ 
&= \sum_{i=1}^{n}i^{-\alpha} - 
\sum_{i=1}^{k-1}i^{-\alpha}-\frac{k^{-\alpha}+n^{-\alpha}}{2}
\end{split}
\end{equation}
so that our truncated Riemann Zeta function is approximated as the sum of the associated integral and of a limited number of terms of the series itself:
\begin{equation}
\sum_{i=1}^{n}i^{-\alpha} \simeq \int_{k}^{n}x^{-\alpha}dx + \sum_{i=1}^{k-1}i^{-\alpha} + \frac{k^{-\alpha}+n^{-\alpha}}{2}.
\end{equation}

\begin{figure}
\begin{center}
\includegraphics[width=0.7\columnwidth]{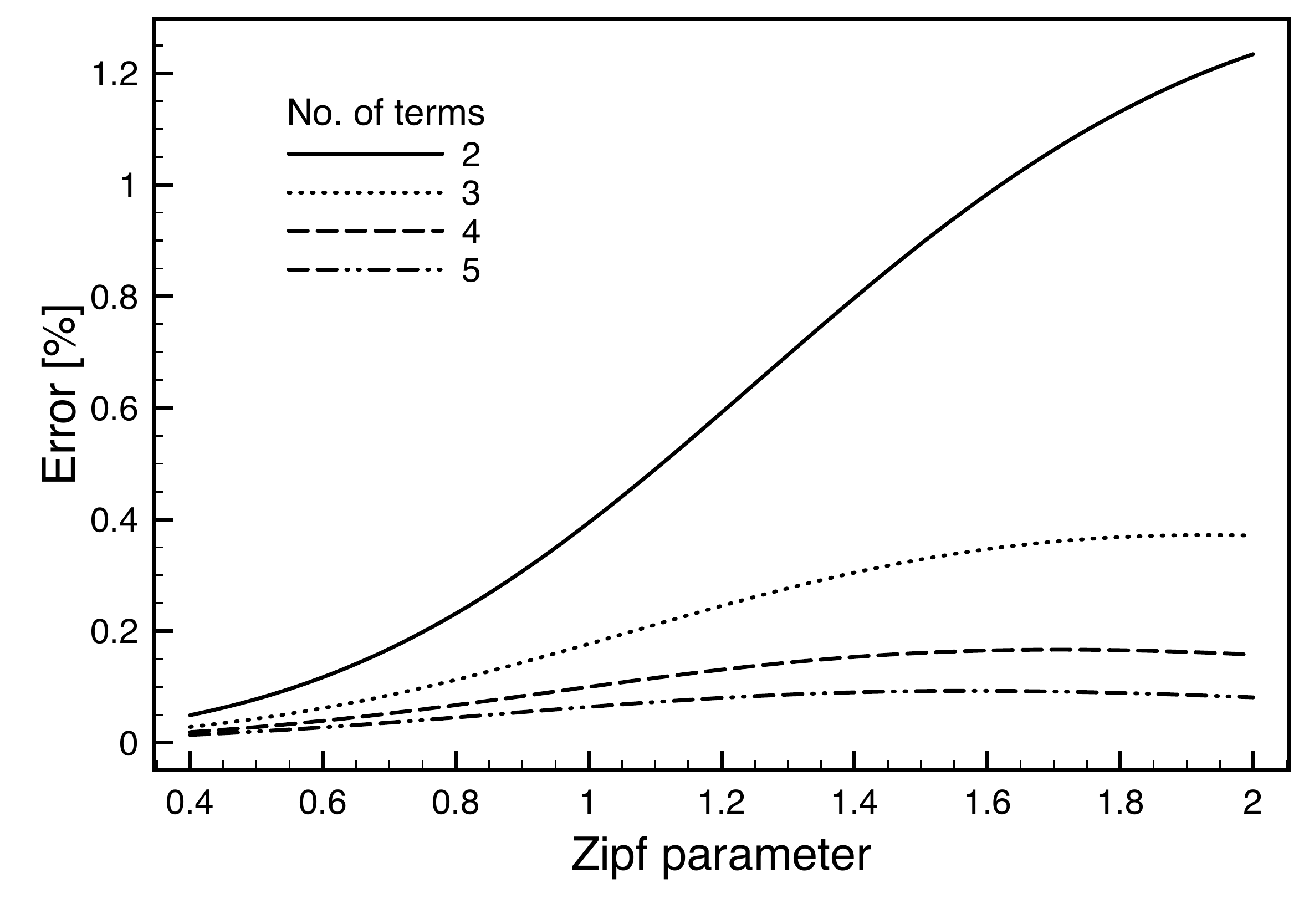} 
\caption{Relative error of the trapezoidal rule approximation (No. of species=100)}
\label{fig:errtrap100}
\end{center}
\end{figure}

\begin{figure}
\begin{center}
\includegraphics[width=0.7\columnwidth]{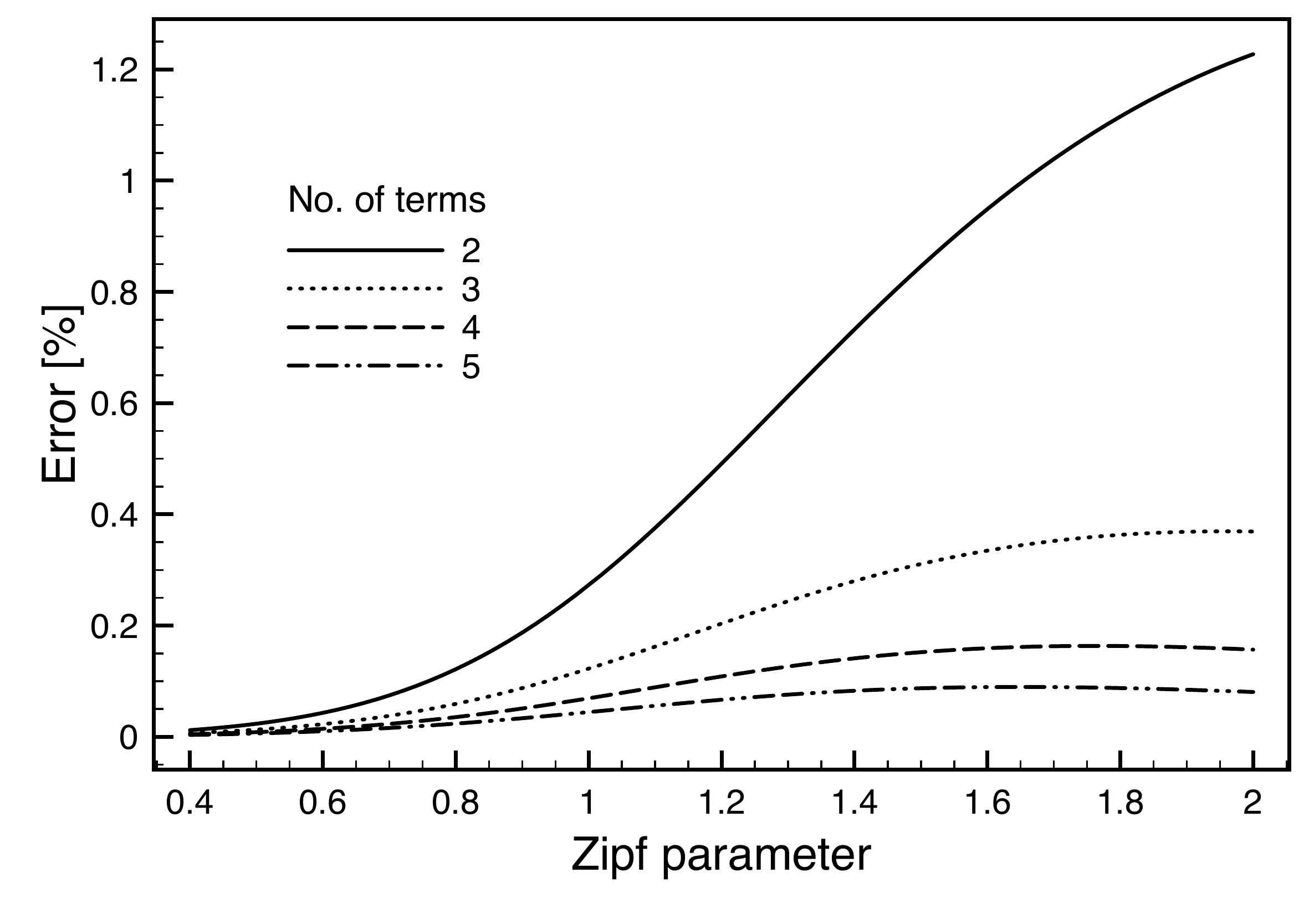} 
\caption{Relative error of the trapezoidal rule approximation (No. of species=1000)}
\label{fig:errtrap1000}
\end{center}
\end{figure}

\begin{figure}
\begin{center}
\includegraphics[width=0.7\columnwidth]{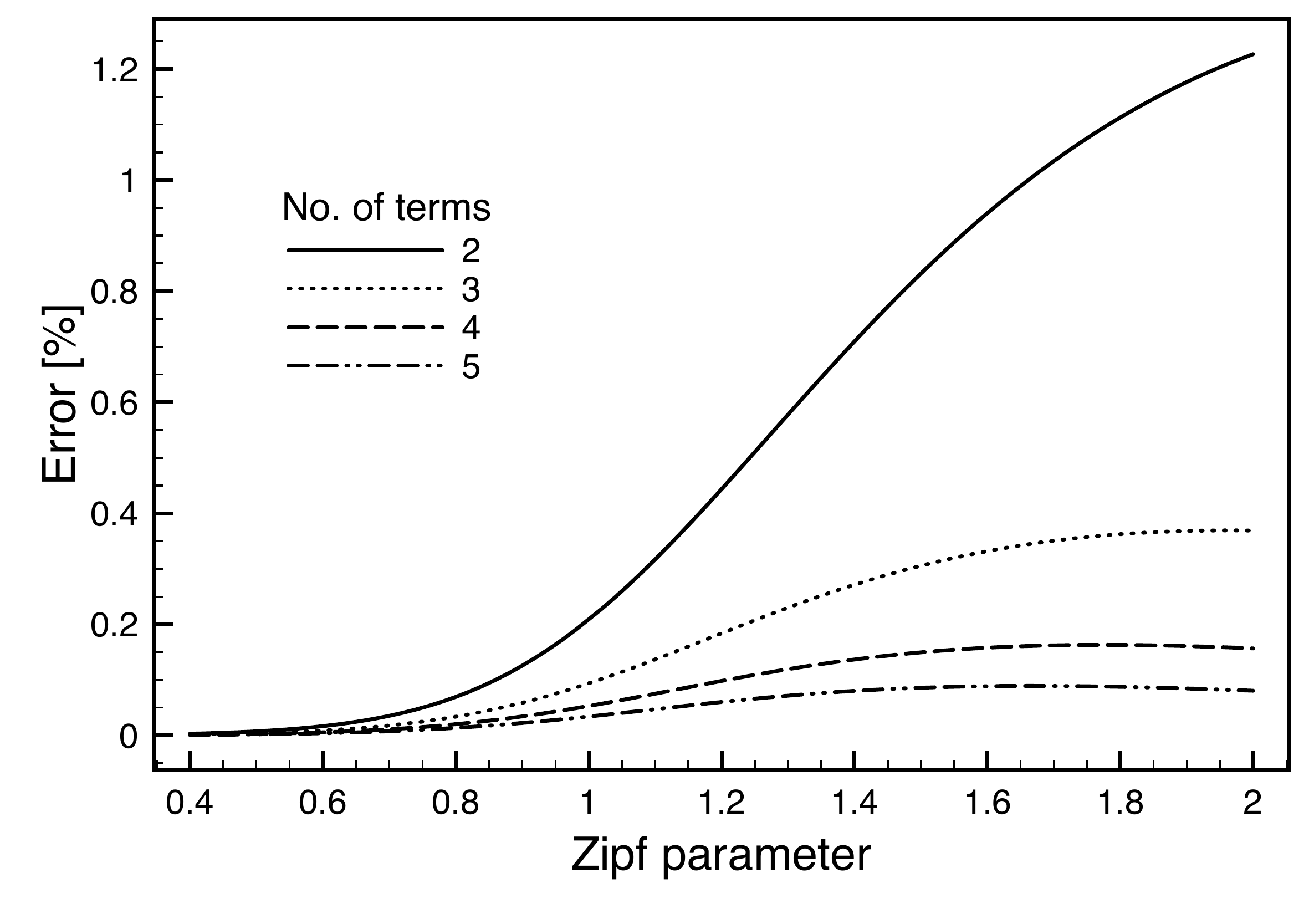} 
\caption{Relative error of the trapezoidal rule approximation (No. of species=10000)}
\label{fig:errtrap10000}
\end{center}
\end{figure}

Recalling the integral appearing in Equation (\ref{appsum}), we have in the end
\begin{equation}
\label{ptrap}
\mathbb{P}[X=r] = \frac{r^{-\alpha}}{n^{-\alpha}\left( \frac{1}{2} + \frac{n}{1-\alpha}\right)+k^{-\alpha}\left( \frac{1}{2}-\frac{k}{1-\alpha}\right)+\sum_{i=1}^{k-1}i^{-\alpha}}.
\end{equation}
In this expression the number of additional terms is set by $k$, for which we can use values as low as 2 or 3. For the simplest case ($k=2$) Equation (\ref{ptrap}) reduces to
\begin{equation}
\label{ptraprid}
\mathbb{P}[X=r] = \frac{r^{-\alpha}}{n^{-\alpha}\left( \frac{1}{2} + \frac{n}{1-\alpha}\right)+2^{-\alpha}\left( \frac{1}{2}-\frac{2}{1-\alpha}\right)+1}.
\end{equation}
In order to evaluate the quality of this approximation we can again use the relative error, which is independent of the rank $r$:
\begin{equation}
\label{err3}
\begin{split}
\epsilon _{\textrm{trap}} &= \frac{\mathbb{P}_{\textrm{trap}}[X=r]-\mathbb{P}[X=r]}{\mathbb{P}[X=r]} \\ &= \frac{\sum_{i=1}^{n}i^{-\alpha}}{n^{-\alpha}\left( \frac{1}{2} + \frac{n}{1-\alpha}\right)+k^{-\alpha}\left( \frac{1}{2}-\frac{k}{1-\alpha}\right)+\sum_{i=1}^{k-1}i^{-\alpha}}-1.
\end{split}
\end{equation}
The error achieved by the trapezoidal approximation is plotted, for three different values of the number of species and four different choices for $k$, in Figures \ref{fig:errtrap100} through \ref{fig:errtrap10000}.

Now we obtain errors that are at most around 1\% for the simplest approximation ($k=2$) and below 0.1\% for a more complicated approximation ($k=5$).
In addition we can note that the error is smaller the larger the number of species.

\section{Conclusions}
Three approximations, based on the use of the associated integral, have been proposed and evaluated for the truncated Zeta distribution function. Of these the approximation based on the trapezoidal rule is the best, achieving a relative error which is typically lower than 1\% on a very wide range of values of the Zipf parameter and is under 0.1\% for its most typical values. 


\end{document}